\documentstyle[twocolumn]{mn}
\input epsf
\def\etal{{\rm et al.\ }}
\def\gsim{ \lower .75ex \hbox{$\sim$} \llap{\raise .27ex \hbox{$>$}} }
\def\lsim{ \lower .75ex \hbox{$\sim$} \llap{\raise .27ex \hbox{$<$}} }
\def\hmin{\ifmmode{h^{-1}}\else{$h^{-1}$}\fi}
\def\kms {\ifmmode{{\rm km\,s}^{-1}}\else{km\,s$^{-1}$}\fi}
\def\msun{\ifmmode{{\rm M}_\odot}\else{${\rm M}_\odot$}\fi}
\def\Mpc{{\rm Mpc}}
\def\hminMpc3{\ifmmode{h^3{\rm Mpc}^{-3}}\else{$h^3{\rm Mpc}^{-3}$}\fi}
\def\figinsert#1#2{\epsfbox{#1}\message{#2} }  

\title{Projection effects in cluster catalogues}

\author[M.P. van Haarlem et al.]
       {M.~P.~van~Haarlem,$^{1,2}$ C.~S.~Frenk,$^1$ S.~D.~M.~White$^3$\\
       $^1$Physics Dept., Univ. of Durham, South Road, Durham DH1 3LE\\
       $^2$NFRA, P.O. Box 2, 7990 AA Dwingeloo, The Netherlands\\
   	 $^3$Max Planck Institut f\" ur Astrophysik, 
	       Karl-Schwarzschild-Stra\ss e 1, 
          D-85740 Garching bei M\" unchen, Germany}

\date{Accepted 1996 December 31;
      Received 1996 December 12;
      in original form 1996 May 8}

\begin{document}

\maketitle

\begin{abstract}
We investigate the importance of projection effects in the
identification of galaxy clusters in 2D galaxy maps and their effect on
the estimation of cluster velocity dispersions. From large N-body
simulations of a standard cold dark matter universe, we construct
volume-limited galaxy catalogues that have similar low-order clustering
properties to those of the observed galaxy distribution. We then select
clusters using criteria tailored to match those employed in the
construction of real cluster catalogues such as Abell's. We find that
our mock Abell cluster catalogues are heavily contaminated and
incomplete. Over one third (34$\pm$6 per cent) of clusters of richness
class R$\geq$1 are miclassifications arising from the projection of one
or more sub-clumps onto an intrinsically poor cluster. Conversely,
32$\pm$5 per cent of intrinsically rich clusters are missed altogether
from the R$\geq$1 catalogues, mostly because of statistical
fluctuations in the background count. Selection by X-ray luminosity
rather than optical richness reduces, but does not completely
eliminate, these problems.  Contamination by unvirialised sub-clumps
near a cluster leads to an overestimation of the cluster velocity
dispersion which can be very substantial even if the analysis is
restricted only to galaxies close to the cluster centre. Thus, the
distribution of cluster masses -- often used to test cosmological
models -- is a highly unreliable statistic. The median value of the
distribution, however, is considerably more robust because the main
effect of contamination is to create an artificial tail of high
velocity dispersion clusters. Improved estimates of the cluster
velocity dispersion distribution require constructing new cluster
catalogues in which clusters are defined according to the number of
galaxies within a radius about three times smaller than the Abell
radius.
\end{abstract}

\begin{keywords}
galaxies: clusters: general -- cosmology: theory	
\end{keywords}

\section{Introduction}
Clusters of galaxies are a major source of cosmological information.
Because of their large luminosity they can be detected, and their
properties can be measured with relative ease, out to large
distances. This makes it possible to exploit their special
characteristics as the most massive nonlinear objects in the
Universe. In hierarchical clustering theories for the formation of
structure, clusters are associated with the rare high peaks in the
primordial density field on scales of a few megaparsecs. As a result,
their mass and abundance are very sensitive to the amplitude of mass
fluctuations on these scales (Frenk \etal 1990; White, Efstathiou \&
Frenk 1993; Viana \& Liddle 1996; Eke, Cole \& Frenk 1996b). The epoch
of cluster formation and the rate at which the cluster population
builds up is, similarly, a strong function of the mean density
parameter, $\Omega$ \cite{laco93,eke96b}, as is their degree of
internal substructure (Richstone, Loeb \& Turner 1992; Mohr \etal 1995;
Wilson, Cole \& Frenk 1996).  The clustering properties of clusters
depend primarily on the shape of the power spectrum of mass
fluctuations and have been a subject of much debate for over 20 years
\cite{bs83,dalton94,eke96a}.  Finally, rich clusters have recently been
used to map the local density field (Plionis \etal 1996, in
preparation).

The use of galaxy clusters as cosmological diagnostics relies on the
availability of statistical samples, selected according to a
well-defined property, such as richness, mass, or X-ray temperature.
Traditionally, the source of such samples has been Abell's (1958)
cluster catalogue. The integrity of the Abell catalogue, however, has
been questioned over the years (e.g. Fesenko 1979a,b; Lucey 1983; Frenk
\etal 1990). Even Abell himself made it quite clear that the
completeness and homogeneity of his catalogue were suspect. Partly to
overcome these shortcomings new cluster catalogues were constructed in
the early 1990s based, as Abell's, on photographic material, but
replacing eye-ball identifications by automated scans of digitized
plates. These procedures have produced the APM \cite{dalton92} and
Edinburgh-Durham cluster catalogues (EDCC; Lumsden \etal 1992).
Computer manipulation of the galaxy images has allowed a degree of
uniformity and repeatability to be reached that was impossible in
Abell's days. Nevertheless, Abell clusters remain the best studied, and
Abell's catalogue the main source from which samples are drawn for
statistical studies.

Whether by eye or by computer, catalogued clusters are identified as
two-dimensional objects, seen against a strongly clustered
background. The enormous column depth to a cluster makes projection
effects inevitable.  Indeed, spectroscopic follow-up of Abell and ACO
clusters \cite{aco} often reveals several clumps of galaxies lined up
in the direction to a rich cluster (see Katgert \etal 1996 for a recent
study of a large sample).  Although these clumps enhance the apparent
richness of the cluster, the dominant concentration along the
line-of-sight is often rich enough to emit X-rays (e.g. Briel \& Henry
1993).  This fact alone, however, says little about the importance of
projection effects or the completeness of optically selected cluster
samples.

Since the abundance of clusters declines very rapidly with richness or
mass, even a small amount of contamination can compromise statistical
studies in which completeness and/or homogeneity are required. The
cluster two-point correlation function is a good example. Values for
the correlation length, $r_0$, differing by almost a factor of two have
been strongly advocated by various workers
\cite{bs83,postm92,efst92,nich92,dalton94}.  According to Bahcall \&
West (1992) the differences are due to the different limiting
richnesses of the various samples but others have claimed that they are
due (at least in part) to a misclassification of poor clusters which
are placed into a higher Abell richness class as a result of
contamination by the halos of rich clusters
\cite{suth88,dekel89,efst92}.  Eke \etal \shortcite{eke96a} have argued
that a combination of different selection procedures and a richness
dependence of the clustering strength also contributes to these
differences.

While even distant sub-clumps artificially enhance the apparent
richness of a cluster, subclustering in its immediate vicinity causes
its velocity dispersion to be overestimated (e.g. Frenk \etal 1990).
This effect is the likely cause of the poor correlation between the
X-ray temperature and velocity dispersion of the most massive clusters
\cite{dajofo95} and it vitiates comparisons between cluster masses
determined from X-ray, optical and gravitational lensing data
\cite{fahl94}.  The distribution of cluster masses or velocity
dispersions has been used as a discriminant of different cosmological
models (e.g. Weinberg \& Cole 1992; Bahcall \& Cen 1993; Lubin \etal
1996).  These comparisons tend to rely heavily on the behaviour of the
high mass end of the distribution which, unfortunately, is particularly
sensitive to contamination due to substructure. Masses derived from
X-ray data are more reliable (Evrard, Metzler \& Navarro 1996; but see
Balland \& Blanchard 1995), but since cluster samples are almost
invariably selected by their optical properties, the inferred
distributions of X-ray properties are also subject to the kind of
uncertainties discussed above.  Furthermore, by virtue of the fact that
optically selected cluster catalogues all have a lower cut-off in
richness, there is an in-built bias in these samples against low mass
clusters.

It seems clear that the use of cluster properties as cosmological
diagnostics requires a detailed understanding of the biases introduced
by projection effects and contamination in cluster catalogues. The aim
of this paper is to set up a methodology for quantifying such biases
using mock galaxy catalogues constructed from N-body simulations. In
this paper we analyze mock catalogues constructed from standard
$\Omega=1$ cold dark matter (CDM) simulations and we concentrate on
Abell clusters. Our procedures, however, can readily be extended to
other cosmologies and other cluster catalogues. We also use our mock
catalogues for testing alternative procedures for defining clusters and
estimating their properties which avoid some of the biases present in
Abell's catalogue.  Our work extends previous analyses by Frenk
\etal\shortcite{fwed} and White (1991,1992) who used similar
techniques. An earlier assessment of the completeness and contamination
of Abell's catalogue, using Monte-Carlo simulations, was carried out by
Lucey \shortcite{lucey83}. He found that between 15 and 25 per cent of
rich Abell clusters have a true membership that is less than half the
number observed. However, his models contained no dynamical
information, and they did not take into account the clustering
properties of galaxies and clusters.

In Section~2 we give technical details of our simulations and our
method for constructing galaxy catalogues from which mock Abell
catalogues are derived. The galaxy catalogues encapsulate the essential
characteristics of the real situation, although we have not tried to
reproduce the observed properties in every detail.  Section~3 presents
an analysis of the completeness of the cluster catalogues. By
identifying groups along the line-of-sight to each cluster, we
determine the properties of the main concentration of galaxies, and
those of clumps projected onto the cluster. The various categories of
contamination we identify are described in Section~4.  In Section~5 we
discuss how projection effects influence estimates of cluster velocity
dispersions derived from radial velocity measurements and illustrate
how careful cluster selection and interloper removal can improve upon
the accuracy of these estimates.  In Sectio~6 we apply a popular
statistical test for substructure to our data and demonstrate its
potential for flagging clusters whose velocity dispersion estimates are
strongly affected by substructure.  In Section~7 we present the
cluster-cluster correlation function, and illustrate how overlapping
clusters can affect the amplitude of this function.  Finally, we
present a summary of our main results in Section~8.

\begin{figure}
\centering
\centerline{\epsfysize=9.5truecm
\figinsert{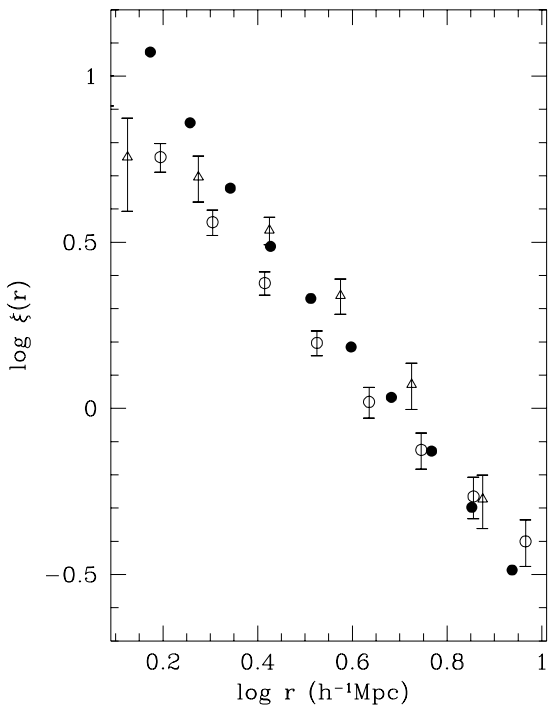}{Figure 1}}
\caption{The real space two-point correlation function of galaxies in
our catalogues compared with observations. The solid circles show
$\xi(r)$ obtained by averaging over 8 independent simulated galaxy
catalogues.  The open circles show Baugh's (1996) deprojection of the
angular correlation function of the APM galaxy catalogue. The triangles
show the real-space correlation function derived by Loveday \etal
(1995) from the Stromlo-APM galaxy redshift survey.}
\end{figure}

\section{Methods}

\subsection{Simulations}

Our analysis employs the set of 8 standard CDM N-body simulations
described by Eke \etal\shortcite{eke96a}.  Each simulation followed
$128^3$ particles in a comoving periodic box 256 \hmin\Mpc \footnote{We
write Hubble's constant as $H_0\, =\, 100\, h\,\kms\Mpc^{-1}$
throughout.} on a side, using Couchman's \shortcite{couchm91} adaptive
P$^3$M code. The mass per particle is therefore $2.2\times
10^{12}\hmin\msun$. The force softening (for an equivalent Plummer
potential) was fixed at $\sim 65\hmin$kpc in comoving coordinates.

Initial conditions were laid down using the CDM transfer function given
by Bardeen \etal \shortcite{bbks} (hereafter BBKS) for adiabatic
fluctuations in a universe with a negligibly small baryon density and
$h=0.5$.  Initial velocities and displacements were computed using the
Zel'dovich approximation as outlined by Efstathiou \etal
\shortcite{edfw} and Davis \etal \shortcite{defw}.  The expansion
factor $a$ was chosen so that $\sigma_8=a$, where $\sigma_8$ is the
$rms$ amplitude of mass fluctuations in a sphere of radius 8
\hmin\Mpc. Each simulation was started at $a$=0.05 and halted at
$a$=0.63, using a time-step $\Delta a$=0.002.  Only the final epoch
$a$=0.63 was used for the present study.  According to White \etal
(1993), Viana \& Liddle (1996) and Eke \etal (1996b) this normalization
is required to obtain approximately the observed abundance of rich
clusters.

\subsection{Constructing Galaxy Catalogues}

It is not yet possible to carry out simulations of cosmological volumes
that are large enough to contain many galaxy clusters and have enough
resolution to model the dissipative processes of star and galaxy
formation. Attempts to introduce galaxies into numerical simulations
are therefore subject to considerable uncertainty. We have implemented
a scheme that produces galaxy catalogues that match the observed
galaxy-galaxy two-point correlation function over the range of
separations of interest and on which we can apply a procedure for
finding clusters that closely mimics Abell's selection criteria. The
prescription we use is based on the peak-background split technique
outlined in BBKS.  Further information about the detailed
implementation may be found in White \etal \shortcite{wfde}.

The BBKS formalism gives the number density of peaks of a certain
height in a field $F$ filtered on a galaxy scale, $r_s$, even though
the simulations do not resolve the field, $F$, on that scale, but only
on a scale $r_b > r_s$.  The scale $r_s$ we associate with individual
galaxies and corresponds to ${\sim}10^{12}\msun$.  Now let $F_s$ denote
the Gaussian random density field, $F$, smoothed with a Gaussian filter
of width $r_s$. We associate a galaxy with a density peak in excess of
$\nu_s\sigma_s$, where $\sigma_s$ is the $rms$ value of the field
$F_s$. Similarly, we define the quantities $F_b$ and $\sigma_b$ for the
same field, $F$, now filtered on a scale $r_b$.  The full expression
for the quantity ${\cal N}_{pk}(\nu_s | \nu_b) d \nu_s$, the number
density of peaks in the field $F_s$ with height in the range
$\nu_s\sigma_s$ to ($\nu_s + d\nu_s$)$\sigma_s$, at points where the
background field has the value $\nu_b\sigma_b$, is given in Appendix E
of BBKS. In order to ensure that the correlation properties of the
galaxies reflect those of the underlying distribution of peaks in $F_s$
we choose a suitable filter function to obtain $F_b$. By using a sharp
low-pass filter in $k$-space, we eliminate the correlations between the
difference field $F_s - F_b$ and $F_b$. The price of this is that
oscillations in the correlation function of the background field only
vanish asymptotically. However, this effect is negligible on the scales
of interest and, since we are not seeking to provide an exact match to
the galaxy distribution, this scheme is quite sufficient for our
purposes.

The procedure outlined above was implemented as follows: the initial
density distribution (sampled on a $128^3$ grid) was smoothed by
removing all power below $r_b$=8.75\hmin\Mpc. A tabulated version of
${\cal N}_{pk}(\nu_s | \nu_b) d \nu_s$ was then used to find the 'peak
number' associated with each point on the grid of the smoothed initial
density field. In order to construct a volume-limited catalogue, we
determined the total number of galaxies by requiring that the
luminosity density of our model catalogue be consistent with recent
determinations of the luminosity function (e.g. Loveday \etal 1992;
Marzke \etal 1994). Our adopted value of $\rho_L = 0.0176 L_*\hminMpc3$
lies in the range bracketed by the observational data which differ by
rather large amounts. This normalization reproduces the observed
abundance of Abell clusters ($\sim 8\times 10^{-6} \hminMpc3$,
Scaramella \etal 1991) when we identify clusters using the method
described in the next subsection.  The only other two parameters
$\nu_s=1.2$ and $r_s=0.54\hmin\Mpc$ were chosen to match the observed
galaxy-galaxy correlation function, $\xi_{gg}$ (see Figure 1). For our
chosen values of $\nu_s$ and $r_b$, the mean luminosity associated with
each peak is 0.87 $L_*$, and the mean luminosity associated with each
particle is 0.14 $L_*$.  (With the large number of particles in these
simulations, no oversampling is necessary, even in the densest regions;
see White \etal 1987).  Where necessary, we have assumed that the
distribution of galaxy luminosities follows a Schechter function:
\begin{equation}
\phi(L)dL = N_* L^{-\alpha} \exp(-L)/\Gamma(2-\alpha)dL,
\end{equation}
where $L$ is expressed in units of $L_*$, and $\Gamma$ is the gamma
function. The final galaxy catalogues are constructed by randomly
selecting particles in the simulation volume and identifying them as
galaxies with a probability directly proportional to the `peak number'
associated with the grid point nearest that particle in the initial
conditions.  The galaxy inherits from the particle both its position
and velocity at later times.

\begin{figure}
\centering
\centerline{\epsfysize=9.5truecm
\figinsert{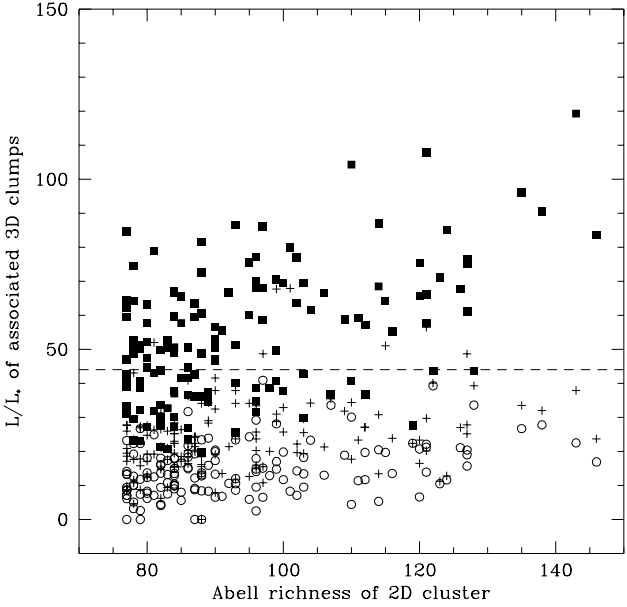}{Figure 2}}
\caption{The total luminosity of the main cluster (solid squares),
together with the second (crosses) and third (open circles) brightest
clumps along the line-of-sight to each cluster, as a function of the
galaxy count in 2D. The dashed line represents the 3D luminosity
threshold of $43 L_*$ required for an Abell cluster of richness class
R=1. No background count has been subtracted from the 2D cluster
richness count.}
\end{figure}

\begin{figure}
\centering
\centerline{\epsfysize=9.5truecm
\figinsert{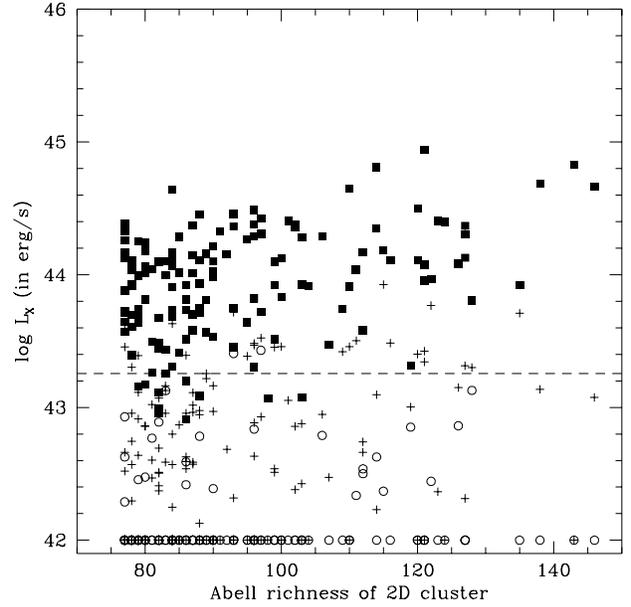}{Figure 3}}
\caption{The X-ray luminosity (within a detect cell of radius
0.75\hmin\Mpc) of the main cluster (solid squares), together with the
second (crosses) and third (open circles) brightest clumps along the
line-of-sight to each cluster, as a function of the galaxy count in 2D.
The dashed line corresponds to a luminosity of 1.8$\times
10^{43}$erg/s, the lowest detection in the survey of Briel \&
Henry~(1993)}
\end{figure}

\subsection{Cluster Selection}
Abell defined a rich cluster as an enhancement of galaxies on the
Palomar Sky Survey plates. To qualify as a cluster, the number of
galaxies within an `Abell radius' ($r_a=1.5\hmin\Mpc$) from the
proposed cluster centre had to exceed a certain number, $n_a$, after
background subtraction.  Galaxies were counted in the magnitude
interval between $m_3$ and $m_3 + 2$, where $m_3$ is the apparent
magnitude of the third brightest galaxy.  For a cluster of Abell
richness class R=1, the lowest richness class that can be considered
reasonably complete, $n_a\geq 50$, while for a cluster of Abell
richness class 2, $n_a\geq 80$.

As starting point for the construction of mock catalogues that
encapsulate the main features of Abell's \shortcite{abell58} cluster
catalogue, we use volume-limited galaxy samples.  We assume that the
luminosity distribution of the cluster galaxies is drawn from a
Schechter function as given in equation (1).  We construct catalogues
that have the correct number of galaxies to cover the interval down to
two magnitudes below that of the median value of the third most
luminous galaxy drawn from a cluster with a true luminosity equal to
that expected for a cluster with R=1.  Since the box length in our
simulations is smaller than the effective path length to a typical
Abell cluster, this limit is equivalent to Abell's counting limit.  The
number of galaxies projected onto a typical model cluster, however, is
smaller than the number of galaxies along the-line-of sight to a
moderately distant Abell cluster. Thus our procedure will underestimate
the degree of contamination by distant background galaxies.

Each volume-limited catalogue was projected along the coordinate axes
onto three orthogonal planes, thus producing three different 2D galaxy
catalogues from each simulation. A friends-of-friends group finding
algorithm (Davis \etal 1985) with a linking length 30 per cent of the
mean inter-galaxy separation was then applied to the 2D galaxy
catalogues. The resulting list of groups is the starting point for the
cluster search. (The final cluster catalogues are insensitive to the
choice of linking length in the range $\sim 0.3-0.6$.) The total
luminosity projected within the Abell radius, $r_a$, is $N_* L_*$. The
normalization is thus fixed by the quantity $N_*$ which we now
calculate.

The cumulative distribution of luminosities is found by integrating the
luminosity function
\begin{equation}
\psi(L)\equiv \int_L^\infty \phi(L') dL' = \Gamma(1-\alpha,L) N_*
\end{equation}
where $\Gamma(\alpha,L)$ is the incomplete gamma function. The
distribution of the $j$th brightest cluster member is given by
\begin{equation}
\phi_j(L) dL =\frac{\psi^{j-1}(L)}{(j-1)!} \exp{[-\psi(L)]}\phi(L)dL,
\end{equation}
\cite{schechter76}. For the third brightest galaxy this gives,
\begin{equation}
\phi_3(L)dL = \frac{1}{2} \psi^2(L) \exp{[-\psi(L)]} \phi(L) dL
\end{equation}
and consequently
\begin{eqnarray}
\psi_3(L)  & \equiv & \int_L^\infty \phi_3(L') dL' \nonumber \\
           & = &      1 - \left( 1 + \psi(L) + \frac{1}{2}\psi^2(L) 
                   \right) \exp{[-\psi(L)].} 
\end{eqnarray}
The median luminosity of the third brightest galaxy, $\overline{L}_3$,
is found by solving $\psi_3(L)= \frac{1}{2}$, which gives
$\psi(\overline{L}_3) = 2.674$ and therefore $\overline{L}_3$ is the
solution to:
\begin{equation}
\psi(\overline{L}_3) = \int\limits_{\overline{L}_3}^\infty \phi(L') dL',
\end{equation}
which yields $\overline{L}_3$=2.11.  Bearing in mind that Abell's
richness count is defined over a 2 magnitude interval, we get the
following relation between the normalization parameter, $N_*$, and the
galaxy count, $n_a$:
\begin{equation}
n_a = \frac{N_*}{\Gamma(2-\alpha)} \int\limits_{0.1585
      \overline{L}_3}^{\overline{L}_3} L^{-\alpha} \exp(-L) dL.
\end{equation}
For R=1 clusters, $N_*=60$ for $\alpha=1$. $N_*$ depends only weakly on
the value of $\alpha$ (it varies by less than 10 per cent over the range
$\alpha$=1.0$-$1.5).  $N_*$ is therefore not very sensitive to the
highly correlated and poorly constrained luminosity function
parameters.

The 50 galaxies between $m_3$ and $m_3 + 2$ represent a total
luminosity within $r_a$ of 60$L_*$. Thus, to ensure that all R$\geq$1
clusters in the simulation volume, $V$, are detected requires a
volume-limited catalogue with
\begin{equation}
N_{gal} = \rho_L \frac{50}{N_*L_*} V
\end{equation}
galaxies. Since the total number of galaxies that resides in the
clusters is small compared with the total number in the box, we assume
that the contamination by foreground and background galaxies is
proportional to the volume projected onto the cluster, i.e. to the
number of galaxies within a cylinder of volume $\pi r_a^2 l_{box}$
($l_{box}=V^{1/3}$) centred on the cluster. Within an Abell radius we
expect, on average, 27 background galaxies in addition to at least 50
cluster members.

Each of the groups identified using the friends-of-friends algorithm in
the 2D galaxy catalogue was checked to see whether the number of
galaxies within the Abell radius exceeded the 76 needed to qualify as
an R=1 cluster. The poorer of a pair of overlapping clusters (projected
separation $< 2 r_a$) was removed. On average the number of clusters
found per catalogue was 139, corresponding to a number density $\sim
8\times 10^{-6} \hminMpc3$.

One of the quantities that we wish to calculate is the fraction of
clusters that would actually meet Abell's criterion if this were
applied in three dimensions. For this we need to convert the 2D
luminosity threshold found above ($N_*=60 L_*$) to a 3D threshold. This
requires making a correction for the galaxies that are projected onto
the cluster, but lie outside $r_a$. We assume that an average spherical
cluster can be adequately described by a power-law density profile
$\rho(r)\propto r^{-\gamma}$.  The correction is the ratio of
$M_p(r_a)$, the mass seen in projection within a cylinder of radius
$r_a$, to the total mass $M(r_a)$ within the aperture in 3D. For the
power-law density profile these masses are,
\begin{equation}
M_p(r)=\frac{4\pi}{3-\gamma} r^{3-\gamma}
\frac{\Gamma(\frac{1}{2})\Gamma(\frac{\gamma-1}{2})}
{2\Gamma(\frac{\gamma}{2})},
\end{equation}
\begin{equation}
M(r)=\frac{4\pi}{3-\gamma} r^{3-\gamma}.
\end{equation}
Taking $\gamma=2.2$, the value derived by Lilje \& Efstathiou
\shortcite{lilefs88} from their determination of the cluster-galaxy
cross-correlation function, we find a ratio $M(r_a)/M_p(r_a)$=0.72 and
therefore a 3D threshold for richness class R$\ge$1 of $43 L_*$.  The
assumed value of $\gamma$ is in good agreement with $\gamma = -2.3 \pm
0.4$ found directly from the luminosity weighted particle distribution
over the interval between 0.1 and 1.5 {\hmin\Mpc}.  The mean slope
found for the 2D cluster catalogue is $-2.0\pm 0.3$, slightly lower as
a result of contamination caused by projection effects.
\begin{figure*}
\centering 
\centerline{
\epsfxsize=18.0truecm 
\epsfbox[0 0 540 392]{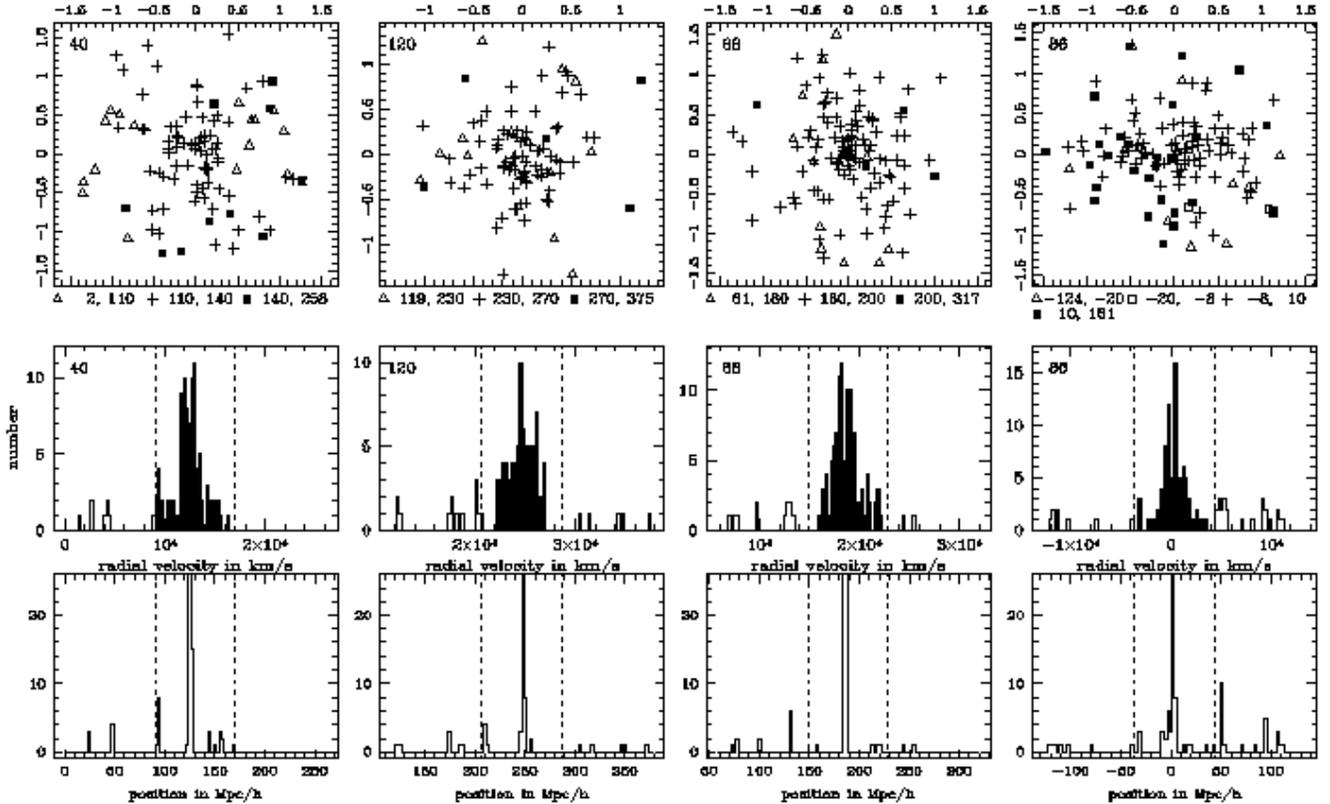}
\message{Figure 4}}
\caption{The top panel shows the distribution of galaxies in two R=1
clusters (numbers 40 and 120, left two columns) and two R=2 clusters
(numbers 68 and 86). Different sections along the line-of-sight have
been plotted with different symbols, as indicated in the legend at the
bottom of each figure. (The numbers in the legend are intervals in
units of Mpc/h.) Open symbols are used for galaxies in front and solid
symbols for galaxies behind the cluster, while crosses denote galaxies
that belong to the cluster. The middle row shows the corresponding
distribution of radial velocities in a cylinder centred on the
cluster. The shaded part of the histogram shows the galaxies that are
classed as cluster members after a $3\sigma$ clipping procedure (see
below); the unshaded regions have been classed as foreground and
background objects. The lower panel shows true positions along the
line-of-sight. Note that these lower two panels are centred on the
cluster. The dashed lines denote an 8000 km/s window, centred on the
cluster, used to make an initial selection for the determination of
velocity dispersions. These clusters are only weakly contaminated by
projection effects. }
\end{figure*}

\section{The Completeness of Cluster Catalogues}
The effects of contamination by foreground and background galaxies on
the completeness of the Abell catalogue were first considered by Lucey
\shortcite{lucey83}.  His Monte-Carlo simulations, however, were crude
because they did not take into account the clustering of the
contaminating galaxies. By contrast, our mock galaxy catalogues
reproduce the low-order clustering statistics of the real galaxy
distribution over the range of scales of interest and so they provide a
much better approximation to the source of projection effects.

To investigate the reality of clusters identified in projection, we
compare our mock 2D cluster catalogues with catalogues of clusters
identified in 3D. The latter were constructed from the centres returned
by the friends-of-friends group finder applied to the dark matter
distribution, using a linking length of 10 per cent of the mean
inter-particle separation. (Again, the results are insensitive to this
choice.)  The luminosity of these 3D clumps was calculated by summing
over the total luminosity associated with the particles contained
within a sphere with radius 1.5\hmin\Mpc.  Only groups with more than
eight particles were considered. With each 2D `Abell' cluster, we
associate all 3D clusters whose centres fall within the Abell radius of
the projected cluster. Figure~2 shows the distribution of luminosities
of the three most luminous 3D groups identified along the line-of-sight
towards each Abell richness class R$\ge$1 cluster in one of our 24
catalogues. The 3D threshold of 43$L_*$ required for richness class R=1
is shown by the dashed line. Averaged over all galaxy catalogues, we
find that 34 $\pm$6 per cent of `Abell' clusters are not associated
with a 3D cluster which, on its own, meets the generalized Abell
criterion. These clusters are only seen above the threshold in
projection because of the superposition of several sub-threshold clumps
along the line-of-sight.  Conversely, we find that 32$\pm$5 per cent of
the 3D clumps brighter than 43$L_*$ fail to be picked out as Abell
clusters in projection. In most cases, such non-detections are due to
fluctuations in the background count as a result of which intrinsically
rich clusters are misclassified as poorer groups. In addition, in 5 per
cent of cases, we find that two 3D clumps above the threshold are
associated with a single 2D object. Thus, in our simulations, about a
third of clusters classified as Abell R$\ge$1 clusters are, in fact,
poorer groups, whilst a similar fraction of intrinsically rich
clusters, are not included in the catalogue at all. We conclude that
catalogues selected using Abell's criteria are neither homogeneous nor
complete to a uniform integrated luminosity limit.

X-ray emission from the hot intra-cluster medium provides an
alternative means of selecting cluster samples. No extensive sample
selected exclusively on the basis of X-ray data exists to date. A first
step in this direction is the catalogue of clusters compiled by Romer
\etal (1994) who correlated a sample of X-ray sources with optical
galaxy counts from digitized photographic plates.  In most studies
however, the cluster catalogue is taken as the starting point for
further selection according to X-ray flux or luminosity (e.g. Nichol,
Briel \& Henry 1994).  Any incompleteness present in the cluster
catalogue is then automatically carried forward to the X-ray sample.
Nevertheless, it is of interest to ask how such X-ray cluster samples
are further affected by projection effects. Our simulations do not
follow the gas component responsible for the X-ray emission from
clusters. However, we can use the results of recent N-body/gasdynamics
simulations of the formation of individual clusters to calculate,
approximately, the expected X-ray luminosity from our clusters. These
simulations show that the collapse and shock heating of a non-radiative
gas during cluster formation establishes a near equilibrium
configuration in which the density profile of the gas closely follows
that of the dark matter (Evrard 1990; Thomas \& Couchman 1992; Navarro,
Frenk \& White 1995, but see Anninos \& Norman 1996).  We can therefore
estimate the expected X-ray emission from a cluster by associating with
each particle in our simulations an `X-ray luminosity' proportional to
the product of the local density and velocity dispersion,
$L_x\propto\rho\sigma_v$.  Both these quantities are calculated by
averaging over the 10 nearest neighbours of each particle. When summed
over a group of particles, the total `X-ray luminosity' has the same
dependence on temperature and density as Bremsstrahlung emission, $L_x
\propto \rho^2 T_x^{0.5}$.

Figure 3 shows the `X-ray luminosities' of the same clusters plotted in
Figure~2. The model luminosities were normalized by setting the median
luminosity of the brightest sub-clump along each line-of-sight equal to
$10^{44}$erg/s, close to the median X-ray luminosity in the 0.5-2.5 keV
band of a sample of 145 high galactic latitude Abell clusters studied
by Briel \& Henry \shortcite{brhe93}. These luminosities were
calculated in an inner sphere of radius 0.75\hmin\Mpc, similar, in
projection, to the optimal detect area recommended by Briel \& Henry
for their sample. The dashed line in the Figure corresponds to a
luminosity of 1.8$\times 10^{43}$erg/s, the lowest 0.5-2.5 keV
rest-frame X-ray luminosity detected by Briel \& Henry. These authors
detected only 46 per cent of the 145 clusters in their sample. In part
this is due to uneven sky coverage which also leads to upper limits on
non-detections that vary between 1.6$\times 10^{43}$ and 2.7$\times
10^{44}$erg/s. These variable limits preclude any conclusions regarding
the completeness of the X-ray samples. Nevertheless, Figure~3 shows
that even X-ray selection is not immune from projection effects. A
significant fraction (16 per cent) of the second-ranked clumps and even
some of the third-ranked clumps along the line-of-sight have X-ray
luminosities above the minimum detected in the data. In a few cases,
the second-ranked clumps have comparable luminosities to the
first-ranked clumps.  Only the brightest clusters, those with $L_X
\gsim 10^{44}$ erg/s, provide a clean, although not necessarily
complete, sample. We therefore conclude that even X-ray selected
samples are, to some extent, contaminated by projection effects,
although this seems to be a weaker effect than for optically-selected
samples. Projection effects are reduced because of the high central
concentration of the X-ray emission and the use of a relatively small
detect area. Examination of an analogous plot to that in Figure~3,
using X-ray luminosities within 1.5\hmin\Mpc\ rather than within 0.75
\hmin\Mpc, shows that projection effects are much stronger for the
larger detect areas.

\begin{figure*}
\centering 
\centerline{
\epsfxsize=18.0truecm 
\epsfbox[0 0 521 371]{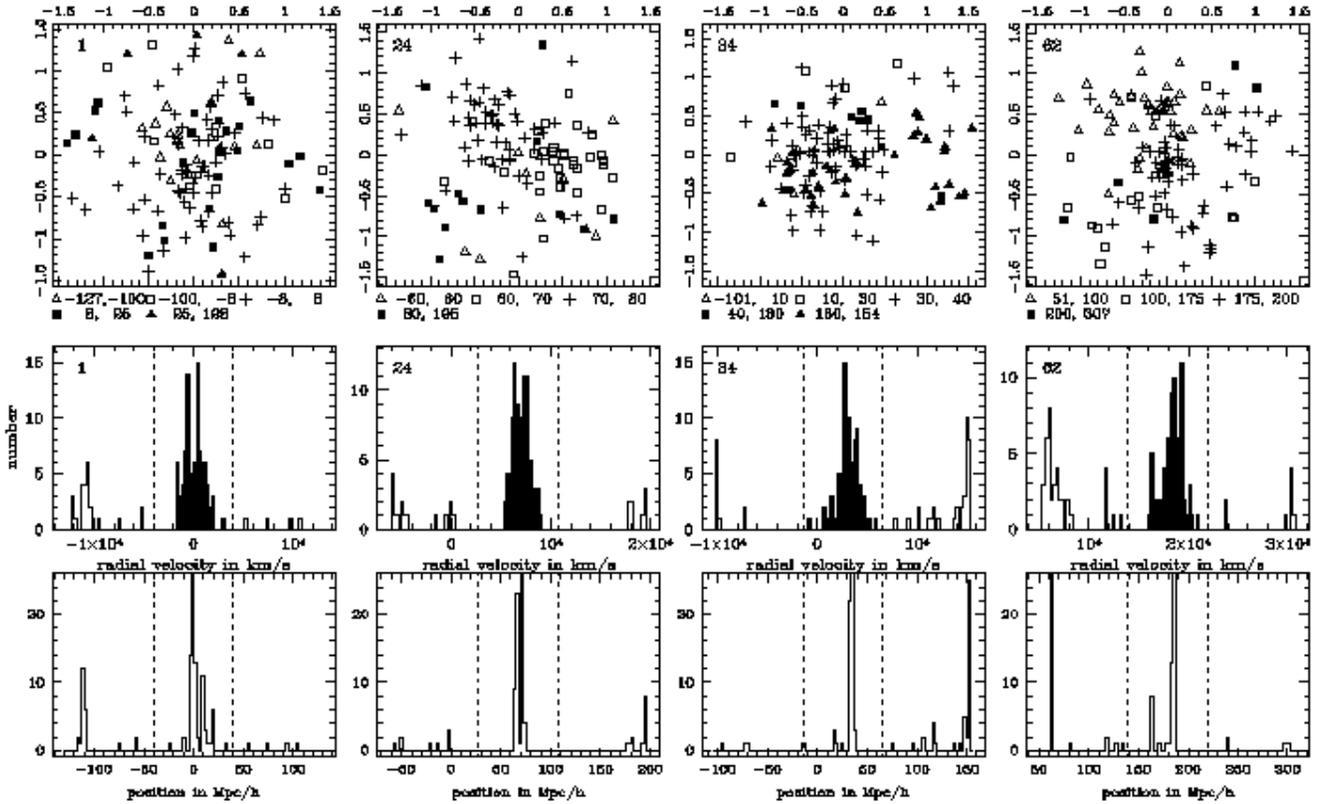}
\message{Figure 5}}
\caption{As Figure~4, but for mildly contaminated clusters. The main
galaxy concentration in these clusters still has richness R$\geq$1
after contamination has been removed. Such clusters clearly belong in
the Abell catalogue, but their richness has been artificially enhanced
by projection effects.}
\end{figure*}

\begin{figure*}
\centering 
\centerline{
\epsfxsize=18.0truecm 
\epsfbox[0 0 537 366]{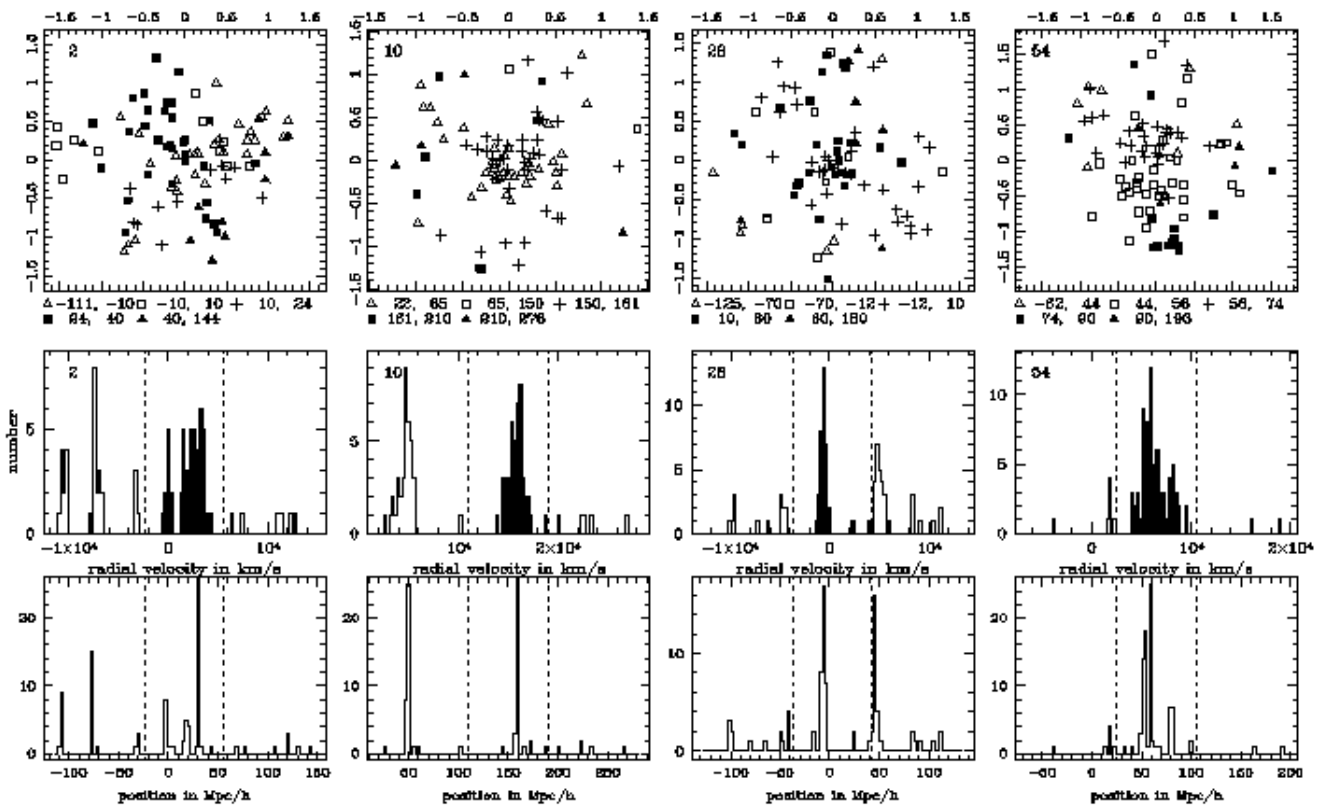}
\message{Figure 6}}
\caption{As Figure~4, but for severely contaminated clusters. The main
galaxy concentration in these clusters has R$<$1 after projection
effects are removed. These clusters should not have been included in
the Abell catalogue.}
\end{figure*}

\begin{figure*}
\centering 
\centerline{
\epsfxsize=18.0truecm 
\epsfbox[0 -80 539 367]{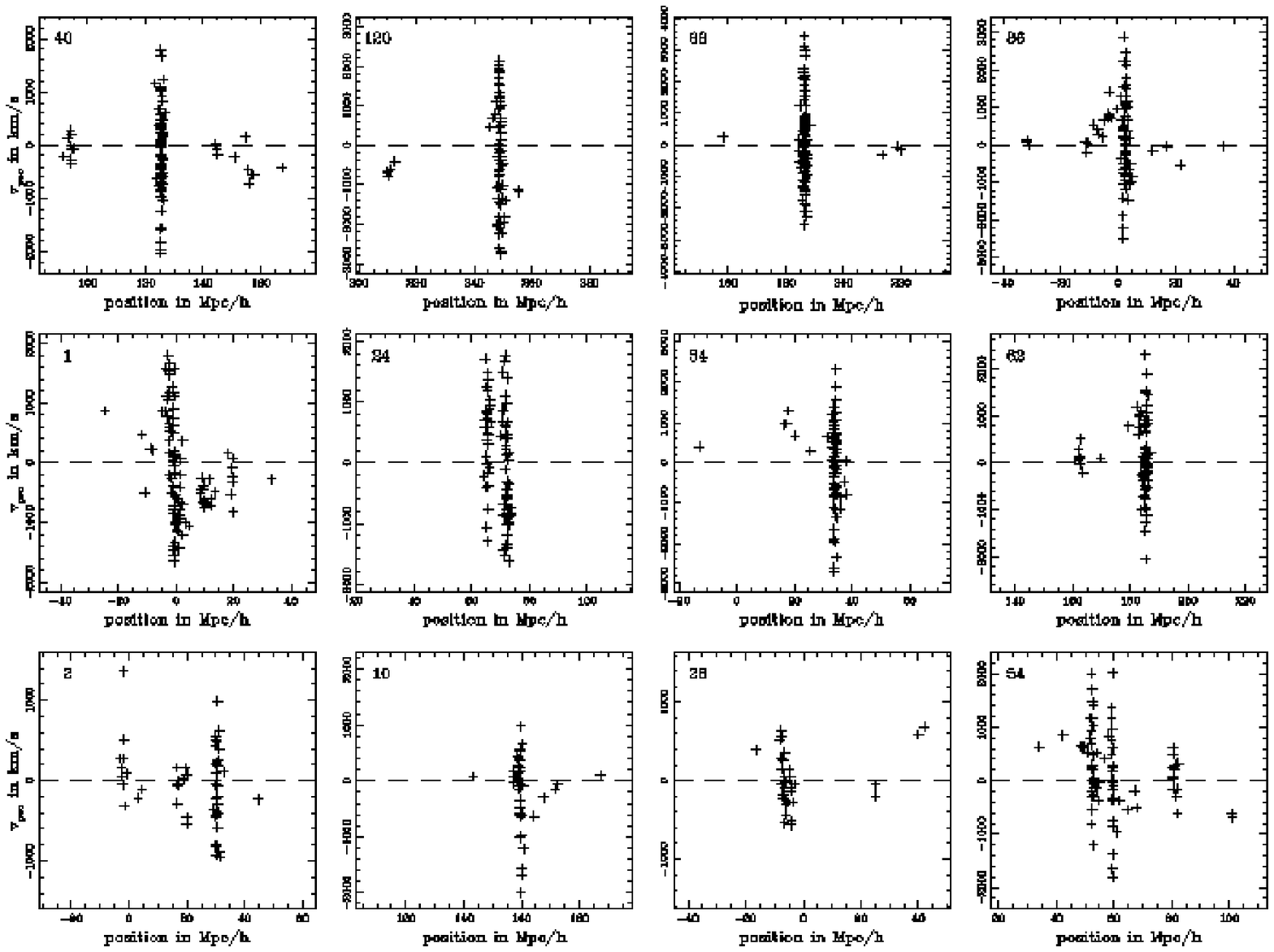}
\message{Figure 7}}
\caption{The radial component of the peculiar velocity plotted against
the line-of-sight position of galaxies in clusters. The top row shows
the R=1 and R=2 clusters of Figure 4, the middle row shows the mildly
contaminated clusters of Figure 5, and the bottom row shows the
severely contaminated clusters of Figure~6.}
\end{figure*}

\section{Structure in the redshift space direction}
The `optically selected' clusters found using the techniques outlined
above show considerable variation in structure and richness. A
significant fraction, $\sim$ 30-40 per cent, of the R$\geq$1 clusters
are almost uncontaminated and clearly meet the criterion derived above
for an Abell cluster in 3D. Examples of such clusters are shown in
Figure~4. The lower panel shows a histogram of the true spatial
distribution of galaxies along the line-of-sight, in bins of width
2.56\hmin{\Mpc} (The vertical axis has been cut off at 26 galaxies.)
These distributions are dominated by a single concentration which is
much richer than any of the sub-clumps along the line-of-sight. The
middle row shows the distribution of radial velocities of all galaxies
in the direction to the cluster. The histograms are not always Gaussian
or even symmetric (Note the high velocity tail at $cz$=20,000\kms\ in
cluster 68), even though the cluster galaxies are well localized in
space. (All cluster members in the lower panel of cluster 68 fall in a
single 2.56\hmin\Mpc\ bin.) The top row shows the distribution of
galaxies on the sky. Different symbols indicate background, foreground
and cluster members (on the basis of the positional information in the
lower row). All clusters appear regular and, in many cases, quite
spherical. Although this is typical of practically all uncontaminated
clusters, there are a few examples of elongated objects (e.g. cluster
86).

The remaining 60-70 per cent of clusters suffer from some kind of
projection effect. Figure~5 shows four examples of clusters that have
been contaminated by other structures. In these cases, had
contamination been removed, the richest remaining group along each
line-of-sight would still have qualified as an R=1 cluster. Although
the velocity histograms do not differ greatly from those of the
uncontaminated clusters in Figure~4, closer inspection reveals that, in
some instances, the foreground and background groups appear to form
identifiable structure when viewed face-on (e.g. the open triangles in
cluster 62); in other cases the contaminating galaxies are distributed
randomly over the entire field (e.g. solid squares in cluster 1).
Cluster 24 is an example of a binary cluster. The smaller component,
containing about 30 galaxies (open squares), is located 5 \hmin\Mpc\ in
front of the main cluster which contains $\sim$55 galaxies. Yet, the
radial velocity distribution in the middle panel is quite symmetric and
shows no clear evidence of bimodality. Although these projection
effects are not as severe as those that lead to poor groups being
detected and classed as rich clusters, they affect the richness
assigned to a cluster and compromise statistical studies including, for
example, estimates of the cluster abundance as a function of richness.

More extreme cases of misidentification and misclassification are shown
in Figure~6. The main peak on which the histograms are centred was
found by filtering with a 4000\,\kms\ top hat function. The main galaxy
concentration fails to meet the 43$L_*$ luminosity criterion for an R=1
Abell cluster in 3D, although the clusters do meet the 2D criterion of
more than 50+27 galaxies in projection. A significant fraction of all
clusters that fall into this category are not as centrally concentrated
as the clusters in Figures~4 and 5. However a surprisingly large number
looks little different, considering that they are pure
superpositions. We find a great variety in the nature of these
superpositions. Some consist of multiple groups of roughly comparable
richness (e.g. clusters 2 and 54), others of two main clumps of
comparable size (e.g. cluster 10).

Closer inspection of the histograms of a random subset of all our
R$\ge1$ clusters reveals that in roughly 14 per cent of cases the main
peak along each line of sight fails to contain even the 30 galaxies
required for an R=0 classification. In 36 per cent of clusters the
count lies between 30 and 50 (R=0), in 40 per cent between 50 and 80
(R=1) and in 10 per cent above 80 (R$\geq$2). An average R=1 cluster
contains 66 per cent of the galaxies that, when projected, form the
cluster in 2D. This is in excellent agreement with our background
estimate.

Although it is often assumed that the radial velocities of virialized
clusters follow a Gaussian distribution, even a small subcluster can
cause a detectable asymmetry. We have examined the degree to which the
radial velocity distribution is Gaussian using a variety of indicators.
In addition to calculating the skewness and kurtosis of the radial
velocity distribution, we have also applied a Lilliefors test (a
Kolmogorov-Smirnov test that takes into account the fact that both the
mean and dispersion are estimated from the dataset itself). In all
cases we estimate confidence intervals by comparing the distribution of
normalized `observed' velocities, $z_i=(v_i-\overline{v})/\sigma$, with
random samples drawn from a normal distribution. All three tests find
that, at the 95 per cent confidence level, between 45 and 55 per cent
of clusters identified in 2D have velocity distributions which are
inconsistent with being Gaussian. The percentage of clusters rejected
at 90 per cent and 99 per cent confidence levels are $\sim 55$ per cent
and $\sim 35$ per cent respectively.

Another illustration of the effects of sub-clumping is given in
Figure~7.  Here we plot the peculiar velocities, $v_{pec}=v_r-v_H$,
where $v_H$ is the Hubble velocity, against position ($v_H/H$) for
galaxies in the twelve clusters shown in Figures~4-6. (In each case we
have limited the total range along the line-of-sight to 100\hmin\Mpc\
for clarity on smaller scales.) Even those clusters which we earlier
regarded as essentially `clean' (top row) suffer from small amounts of
contamination.  The fraction of clusters that does not have a group of
at least 5 members projected onto its face is below 2 per cent. In a
number of cases the mean of the peculiar velocity distribution is
noticeably displaced from zero, as a result of motions induced by
distant matter or by the proximity of a massive neighbouring
cluster. This is clearly seen in the case of the `binary' cluster
(number~24) and in the case of the multiple cluster (number~54).
Several examples show infalling sub-clumps.  Just in front of the main
mass concentration in cluster 86, a group of 15 galaxies is falling
into the cluster, with velocities that increase towards the cluster
core. A similar situation occurs in cluster~1, but now there are at
least two sub-clumps falling in from different directions. In cases
where the cluster fails to meet the criterion for inclusion in the 3D
`Abell catalogue', the number of different small subclusters or groups
that make up the 2D cluster can be substantial (e.g.  cluster~54).

\begin{figure*}
\centering 
\centerline{
\epsfysize=12.0truecm 
\epsfbox[50 200 560 700]{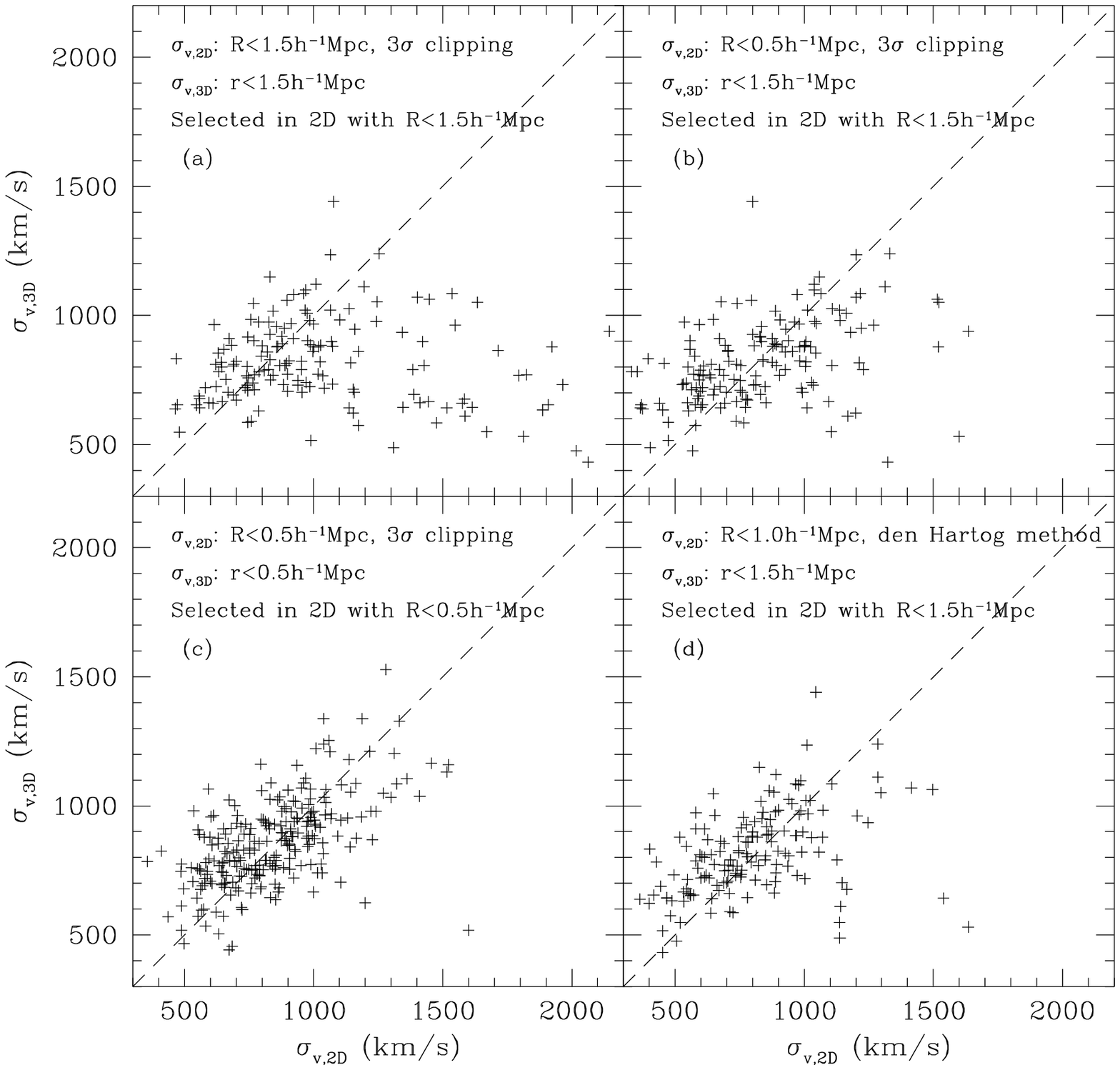}
\message{Figure 8}}
\caption{Comparison of `true' velocity dispersions ($\sigma_{v,3D}$)
and various `observational' estimates ($\sigma_{v,2D}$). In all cases,
the `true' values are $1/\protect\sqrt{3}$ times the 3D velocity
dispersion of the dark matter particles within a sphere of the
appropriate radius. (a) Clusters identified using Abell's criterion and
$\sigma_{v,2D}$ estimated using an optimistic 3$\sigma$ clipping
procedure applied to all galaxies projected within the Abell
radius. (b) Clusters identified using Abell's criterion and
$\sigma_{v,2D}$ estimated using an optimistic 3$\sigma$ clipping
procedure applied only to galaxies projected within a circle of
0.5\hmin\Mpc\ radius. (c) Clusters identified using a scaled version of
Abell's criterion to a radius of 0.5\hmin\Mpc\ and $\sigma_{v,2D}$
estimated using an optimistic 3$\sigma$ clipping procedure on the
galaxies projected within this radius. (d) Clusters identified using
Abell's criterion and $\sigma_{v,2D}$ estimated using the interloper
removal algorithm of den~Hartog \& Katgert (1996).}
\end{figure*}

\section{Cluster velocity dispersions}
The velocity dispersions of galaxy clusters, or the cluster masses
derived from them, are an important cosmological diagnostic. They have
been used, for example, to test specific cosmological models (Weinberg
\& Cole 1992; Bahcall \& Cen 1993; Lubin \etal 1996), and to estimate
the amplitude of mass fluctuations on cluster scales
\cite{evhen91,wef,eke96b}. The distribution of velocity dispersion is
often used in cumulative form, $n(>\sigma_{v,los})$, the number density
of clusters with line-of-sight velocity dispersion greater than
$\sigma_{v,los}$. A closely related quantity is the cumulative X-ray
temperature function, $n(>T_X)$, derived from X-ray observations in the
central parts of clusters.

Frenk \etal \shortcite{fwed} calculated $n(>\sigma_{v,los})$ for an
ensemble of $\Omega = 1$ CDM simulations and Weinberg \& Cole
\shortcite{weincole92} calculated this quantity for various Gaussian
and non-Gaussian models with $\Omega$=1 and $\Omega$=0.2. On the
observational side, Mazure \etal \,\shortcite{mazure96} have recently
estimated $n(>\sigma_{v,los})$ for a sample of 80 clusters from the ACO
catalogue \cite{aco}. They determined each $\sigma_{v,los}$ using at
least 10 (and in 48 out of 80 clusters more than 30) radial velocities
per cluster.  These data represent a considerable improvement upon
previous compilations (e.g. Zabludoff \etal 1993; Bahcall \& Cen 1993).

The main concern when estimating cluster velocity dispersions is the
effect of contamination by unvirialized sub-clumps whose coherent
motion remains hidden by the amplitude of the peculiar velocities in
the cluster region.  As Figures~4-6 show, it is not uncommon for a
superposition of clumps, separated by 10 Mpc or more, to produce an
apparently symmetric and featureless radial velocity
distribution. Using our mock catalogues, we now examine the accuracy of
various conventional methods for estimating velocity dispersions in
`observational' (from now on referred to as 2D) cluster samples by
comparing results obtained from them with the `true' velocity
dispersions computed directly using the full phase-space information
for the dark matter particles within a spherical aperture.

To estimate the velocity dispersion of our 2D samples, we first
convolved the `observed' velocity histogram with a 4000\,\kms\ top hat
filter in order to reject obvious interlopers and obtain an initial
estimate of the mean radial velocity of the cluster. All galaxies with
relative velocities greater than 4000\,\kms\ from the peak of the
convolved histogram were then excluded from the sample. Next, we
applied a standard optimistic $3\sigma$-clipping procedure
\cite{yahvid77} which consists of the following steps: (i) estimate the
mean radial velocity $\overline{v}$ and velocity dispersion $\sigma$;
(ii) delete all galaxies with radial velocity greater than $3\sigma$
away from $\overline{v}$; (iii) estimate $\overline{v}$ and $\sigma$
for the culled sample; and (iv) remove the most extreme galaxy if its
radial velocity is greater than $3\sigma$ from $\overline{v}$. Steps
(iii) and (iv) are repeated until the number of galaxies
stabilizes. This procedure returns a robust and stable estimate of the
cluster velocity dispersion, $\sigma_{v,2D}$.

In panel (a) of Figure~8 we plot the values of $\sigma_{v,2D}$ obtained
by applying this method to the clusters in one of our mock Abell
catalogues against the true values, $\sigma_{v,3D}$, obtained using all
the dark matter particles within the Abell radius, $r_a$, of each
cluster. (The latter are $1/\sqrt{3}$ times the full 3D velocity
dispersion of the dark matter particles.) All qualifying galaxies
projected within $r_a$ were used in the estimate of
$\sigma_{v,2D}$. There is a very poor correlation between these
estimates and the true dispersions. In particular, the 2D distribution
has a tail of high dispersion clusters which is completely
spurious. The true distribution contains only 1 cluster with
$\sigma_{v,3D}>1250$\,\kms\ and none with
$\sigma_{v,3D}>1500$\,\kms. Yet the `observational' sample has 33/153
clusters with $\sigma_{v,2D}>1250$\,\kms and 20/153 with
$\sigma_{v,2D}>1500$\,\kms. These artificially large dispersions are
caused by unvirialized clumps of galaxies projected onto the main
cluster. A good example is cluster 40, shown in Figures~4 and~7. The
derived dispersion for this cluster is $\sigma_{v,2D}=1400$\,\kms, much
larger than the true value $\sigma_{v,3D}=1020$\,\kms. The main culprit
is a group of 8 galaxies located 35\,\hmin\Mpc\ in front of the cluster
which is not eliminated by the optimistic 3$\sigma$ clipping procedure.
According to a Lilliefors test, the hypothesis that the radial velocity
distribution of this cluster is Gaussian cannot be rejected with more
than 25 per cent confidence.

In practice, velocity dispersions for real clusters are most commonly
estimated using only central galaxies, rather than galaxies spread out
over the entire Abell circle, as we have assumed in panel (a) of
Figure~8. In panel (b) of this figure we show the result of estimating
$\sigma_{v,2D}$ from galaxies projected only onto the inner
0.5\,\hmin\Mpc\ of the cluster. By sampling a smaller area, the number
of contaminating clumps is reduced and, as a result, the spurious tail
of high $\sigma_{v,2D}$ clusters is diminished but not altogether
eliminated. The correlation between $\sigma_{v,2D}$ and $\sigma_{v,3D}$
remains rather poor. As before, dispersions for a substantial fraction
of the population are overestimated and, in several cases, they are
significantly underestimated.

The main reason why restricting the radial velocity sample to the inner
regions of the cluster does not produce a more satisfactory result is
that the criteria used to select clusters in the first place already
produces a heavily contaminated catalogue. This is due to the large
search radius employed by Abell. To illustrate this point we construct
a new set of cluster catalogues in which the search radius has been
reduced by a factor of three, to 0.5\,\hmin\Mpc. Assuming a mean radial
density profile, $\rho(r)\propto r^{-2.2}$, the number of galaxies
required for richness class R$\ge$1 within this reduced radius is
24. This scaled Abell criterion produces different cluster catalogues
than Abell's standard criterion. We now apply the same optimistic
$3\sigma$ clipping procedure as before (using galaxies within
0.5\,\hmin\Mpc). The resulting values of $\sigma_{v,2D}$ are compared
with the true values, $\sigma_{v,3D}$ (now calculated using dark matter
particles within a sphere of radius 0.5\,\hmin\Mpc) in panel~(c) of
Figure~8. The result is considerably better. Virtually all the spurious
large dispersions are eliminated and the correlation between the 2D
estimates and the true values is considerably tighter.

The dramatic improvement shown in panel~(c) cannot, of course, be
achieved in practice without replacing Abell's catalogue by a different
one constructed using a smaller search radius. The APM cluster
catalogue \cite{dalton92} fulfills this criterion although the search
radius of 0.75\,\hmin\Mpc\, is somewhat larger than our recommended
value of 0.5\,\hmin\Mpc\ because the latter did not yield a sufficient
number of galaxies in the APM data. However, a weighting scheme was
applied which reduces the weight of galaxies in the outer
0.25\,\hmin\Mpc\ ring. Our mock catalogues indicate that the velocity
dispersion distribution of APM clusters should be considerably more
reliable than that of Abell clusters. Unfortunately an extensive
redshift survey of APM clusters has yet to be undertaken.

Finally, we have tested the interloper removal method proposed by
den~Hartog \& Katgert \shortcite{hartog96}. This method attempts to
find an acceptable range for the radial velocities to be included in
the galaxy sample depending upon the projected distance from the
cluster centre. Using an iterative scheme, the method rejects galaxies
with radial velocities differing by more than $v_{max}(R)$ from the
systemic velocity, where $v_{max}(R)$ is the maximum of the
line-of-sight component of two velocities: (a) the infall velocity for
all positions along the line-of-sight at a given projected separation,
{\it R}, from the cluster centre, and (b) the circular velocity. In
order to calculate the infall velocity a value of $\Omega$ must be
assumed, but den~Hartog \& Katgert \shortcite{hartog96} found that
their results do not depend significantly on the choice of
$\Omega$. This method succeeds in eliminating more galaxies than the
optimistic 3$\sigma$ clipping routine. The comparison between the
velocity dispersions obtained using this method and $\sigma_{v,3D}$ is
shown in panel (d) of Figure~8. The `observational' estimates correlate
better with the true values than the estimates used in Figures~8(a)
and~8(b), but not as well as those in Figure~8(c).  The
den~Hartog-Katgert method tends to underestimate most velocity
dispersions by $\sim 100$ \kms while, at the same time, failing to
completely remove the spurious tail of large dispersions.

\begin{figure*}
\centering 
\centerline{
\epsfxsize=10.0truecm 
\epsfbox[50 200 560 700]{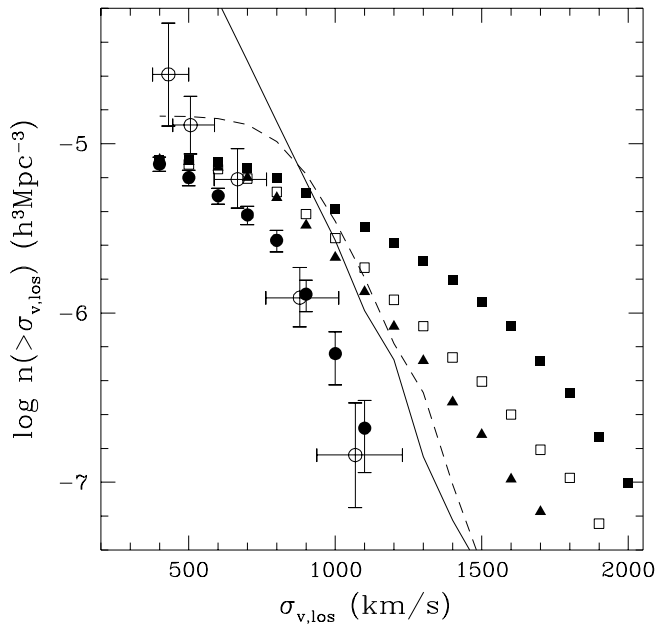}
\message{Figure 9}}
\caption{The cumulative distribution of cluster velocity dispersions
based on all our synthetic cluster catalogues. The solid squares show
velocity dispersions derived from all galaxies projected within $r_a$
from the cluster centre and the open squares from galaxies within
$r_a/3$ (both using $3\sigma$ clipping to remove interlopers). The
triangles were computed from our catalogues using the den~Hartog \&
Katgert interloper removal algorithm. The same algorithm was applied by
Mazure \etal (1996) to a sample of 80 ACO clusters and is shown here as
the solid circles. The open circles were derived from Bahcall \& Cen's
(1993) mass function.  The dashed line shows $n(>\sigma_{v,2D})$ for
clusters selected in 2D using an Abell criterion scaled to $r_a/3$.
The ``true'' distribution, calculated from the velocity dispersion of
the dark matter within $1.5\hmin\Mpc$ is shown by the solid line.}
\end{figure*}

The cumulative distributions of velocity dispersion returned by the
various methods discussed above are compared with the true distribution
in Figure~9. The former were constructed by combining data from all 24
mock cluster catalogues. The true distribution, calculated from the
velocity dispersion of the dark matter particles within 1.5\hmin\Mpc\
of the centre, is shown as the solid line. (All clusters with
$\sigma_{v,los}>500$\,\kms\ are included.) These data agree well with
the cumulative mass distribution of clusters calculated by White \etal
(1993) from N-body simulations of the same model. The different symbols
show the distributions derived for the mock Abell cluster samples using
different methods: the solid squares are the estimates obtained by
applying the optimistic $3\sigma$ clipping to all galaxies within
$r_a$; the open squares correspond to the case when this estimator is
applied only to galaxies within $r_a/3$; the solid triangles give the
result of using the den~Hartog-Katgert method. The dashed line shows
the distribution of dispersions for clusters identified in 2D but using
a reduced search radius of $r_a/3$ and the scaled Abell richness
criterion. The dispersions in this case were also derived using the
3$\sigma$ clipping procedure.  (Note that the total number of clusters
identified in this case is about a factor of 2 larger than the number
of Abell clusters.)

At values of the velocity dispersion below $\sim 850$ \kms\ all the
distributions from the mock catalogues in Figure~9 include only a
steadily decreasing fraction of all the clusters present in the
simulation. This turnover simply reflects the threshold richness
required for selection which biases the sample against low-$\sigma_v$
clusters.  At large values of the velocity dispersion, only the
distribution for clusters identified with a small search radius
approximates the true distribution. All other catalogues overestimate
the number of high dispersion clusters by large factors. Even the
den~Hartog-Katgert method which slightly underestimates the dispersion
of small clusters over-predicts the abundance of clusters with
$\sigma_{v,los}>1500$\,\kms\ by a factor of $\sim 5$. Note, however,
that the tail of the distribution is exaggerated in a logarithmic plot
like this. In fact, at the typical abundance of Abell R$\ge$1 clusters,
all methods perform quite well. For example, the velocity dispersion at
an abundance of $4\times 10^{-6}$ \hminMpc3, half the abundance of
R$\ge$1 clusters, the dispersions obtained from the different methods
lie within $\sim 150$\,\kms\ of the true value. Thus, while the median
velocity dispersion of the Abell cluster population is reasonably well
determined, the high dispersion tail of the distribution is extremely
uncertain.  Figure~9 indicates that statistical studies based on the
median mass or velocity dispersion of the Abell cluster population
(e.g. White \etal 1993) are robust whereas studies that concentrate on
the tail of the distribution (e.g. Bahcall \& Cen 1993) are unreliable.
The APM and EDCC catalogues have a higher number density and are less
affected by projection effects than R$\geq$1 Abell clusters.  One would
therefore expect a sample of such clusters to show a reduction in the
spurious high-$\sigma_v$ tail compared with a sample of Abell clusters

Also plotted in Figure~9 are two determinations of $n(>\sigma_{v,los})$
from observational samples. The solid circles show the estimate by
Mazure \etal\,\shortcite{mazure96} for a sample of 80 clusters analyzed
with the den~Hartog-Katgert algorithm. The open circles show an
estimate derived from Bahcall \& Cen's (1993) mass function of
clusters, assuming that the transformation between mass and velocity
dispersion is that given by their equation~(3) for an isotropic
distribution of galaxy velocities.  The two observational estimates lie
below the model predictions. This is consistent with the results of
Frenk \etal (1990), White \etal (1993) and Eke \etal (1996b) which show
that the abundance of clusters is correctly reproduced in a CDM model
only if $b\simeq 2$.

\begin{figure*}
\centering 
\centerline{
\epsfxsize=16.0truecm 
\epsfbox[0 -80 556 368]{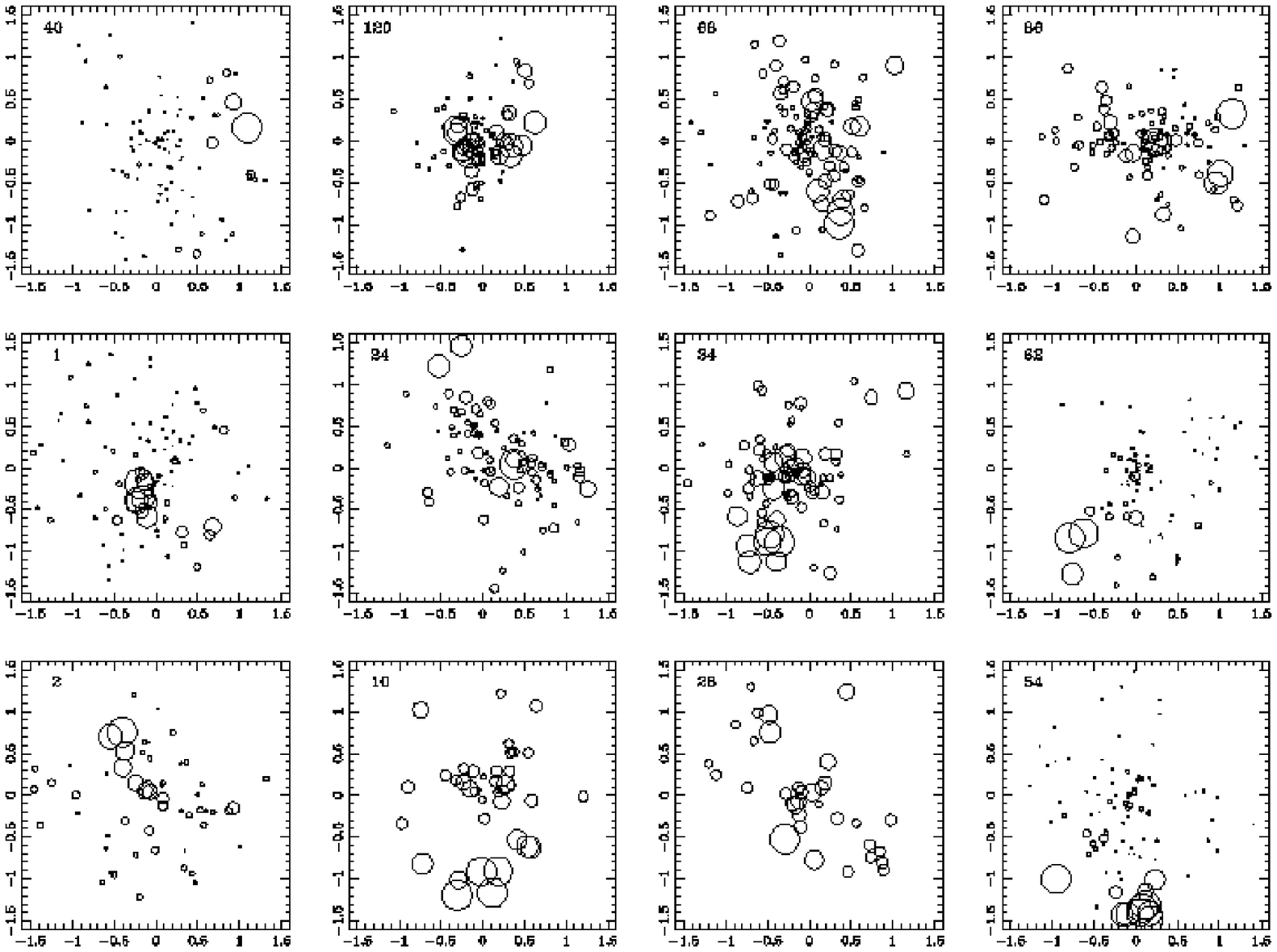}
\message{Figure 10}}
\caption{The Dressler-Schectman test applied to mock clusters. The
symbols show the projected positions of galaxies in our mock
catalogue. The radius of each circle is proportional to
$\exp{\delta_i}$, where $\delta_i$ is a measure of the local deviation
of the radial velocity and velocity dispersion from the average values
for the cluster as a whole.  Foreground and background galaxies have
been removed prior to applying the test using a $3\sigma$-clipping
procedure.}
\end{figure*}

\begin{figure*}
\centering 
\centerline{
\epsfxsize=16.0truecm 
\epsfbox[50 275 575 550]{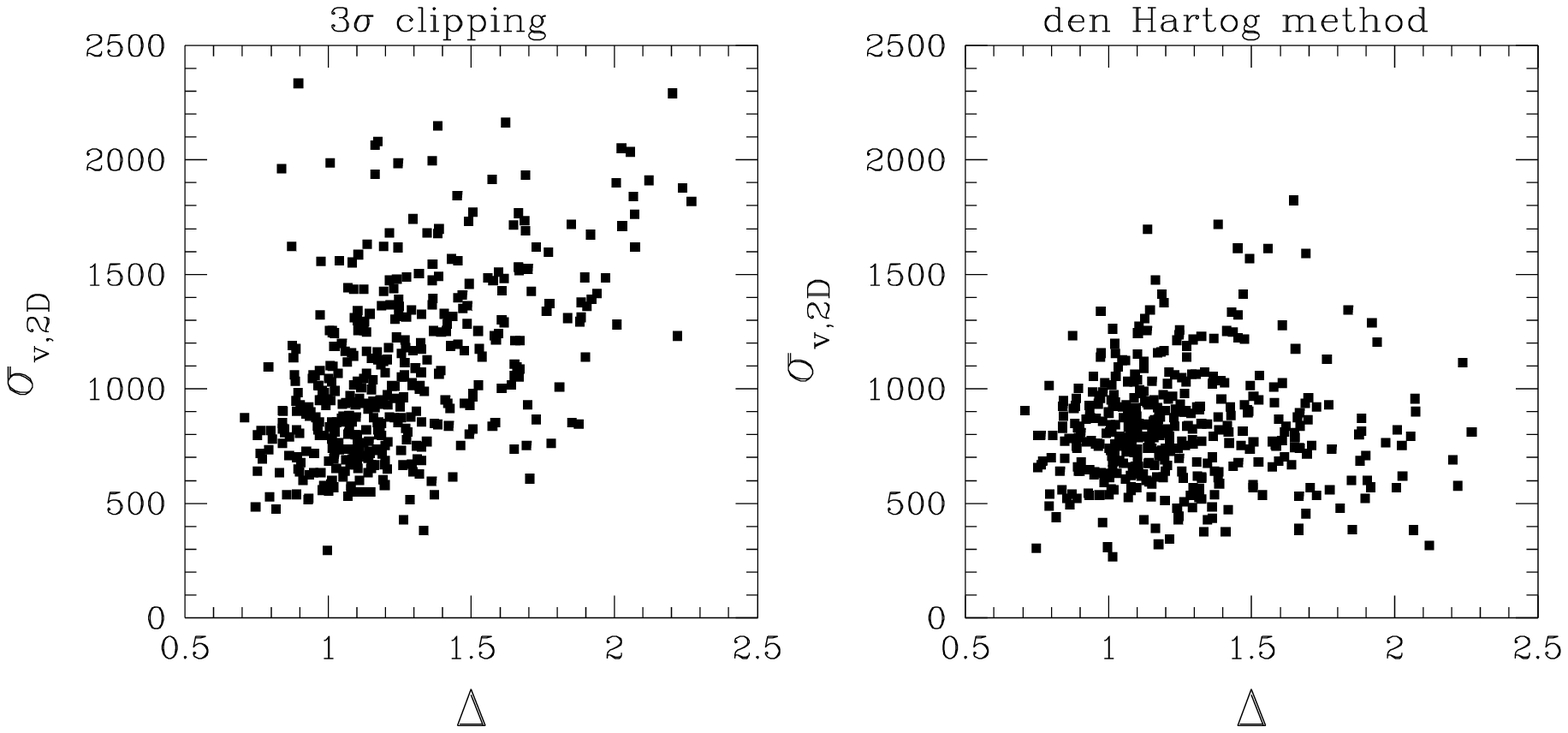}
\message{Figure 11}}
\caption{The correlation between velocity dispersion estimates and the
strength of the substructure signal as measured by Dressler \&
Schectman's $\Delta$ parameter. In the left panel, foreground and
background galaxies were removed using an iterative $3\sigma$ clipping
procedure. On the right, the den~Hartog-Katgert algorithm (see text for
details) was applied to galaxies with a projected separation of less
than 1\hmin\Mpc\ from the cluster centre.}
\end{figure*}

\section{Substructure}
Substructure in a cluster is symptomatic of a recent merger or
accretion event. Typical survival times of accreted groups and
subclusters are comparable to the crossing time (a few times $10^9$ yr)
and thus shorter than the age of the main cluster (Evrard 1990,
Gonzalez-Casado, Mamon \& Salvador-Sole 1994). The presence of
substructure is therefore related to the recent history of a cluster
and provides a useful way of constraining cosmological models in which
cluster growth rates differ (Richstone \etal 1992; van Haarlem \& van
de Weygaert 1993; Bartelmann \etal 1993; Mohr \etal 1995; Wilson \etal
1996). In practice, however, identifying substructure in a cluster
using the properties of its galaxy distribution is difficult, and the
results depend upon the test used to define and detect substructure.

Many statistical tests have been proposed over the years to measure
substructure. The first attempts employed only 2D-data. For example,
Geller \& Beers \shortcite{gelbee82} made contour plots based on
Dressler's \shortcite{dress80} measurements of galaxy positions on the
sky and concluded that $\sim 40$ per cent of clusters showed signs of
substructure. This claim was contradicted by West \& Bothun
\shortcite{wesbot90} and by Rhee, van Haarlem \& Katgert
\shortcite{rvhk91} who applied a range of statistical tests but found
little evidence for substructure in the projected galaxy distribution
in the inner regions of clusters. The inclusion of radial velocity data
allowed more powerful tests to be developed using all observable
properties of the cluster phase-space distribution. Dressler \&
Shectman \shortcite{dreshe88} applied the test that will be described
below to a sample of 15 clusters with an average of 73 measured
redshifts per cluster. They found evidence for substructure in $\sim
40$ per cent of cases. Bird \shortcite{bird94} has claimed that as many
as 85 per cent of well-studied clusters show some evidence for
substructure.  To some extent these disparate results are due to
different definitions of substructure. Nevertheless, the fact that
substructure has been detected even in the Coma cluster (Fitchett \&
Webster 1988; Mellier \etal 1988; White, Briel \& Henry 1993, Colless
\& Dunn 1995, Biviano \etal 1996), usually regarded as the archetypal
relaxed rich cluster, indicates that inhomogeneities are more prevalent
than at first thought.

We can use our mock cluster catalogues to assess some of the most
widely used methods for characterising substructure and to test ways in
which these methods may be used to improve cluster mass determinations
from optical data. It should be noted that our detailed quantitative
results will depend on our assumed cosmological model, particularly on
our adopted value of $\Omega=1$. This choice leads to more prevalent
substructure than models with a low value of $\Omega$ (Richstone \etal
1992, Mohr \etal 1995, Wilson \etal 1996).  Specifically, we consider
here the algorithm proposed by Dressler \& Shectman (1988) which
attempts to identify dynamically distinct subunits within the
cluster. This test is sensitive to local deviations of the mean
velocity and velocity dispersion relative to the global values
determined for the cluster as a whole. For each galaxy in the sample, a
local mean velocity $\overline{v}_{local}$ and a local velocity
dispersion $\sigma_{local}$ are computed using the radial velocities of
the galaxy itself and of its 10 nearest projected neighbours.  The
combined deviations are given by the quantity
\begin{equation}
\delta_i^2 = \frac{11}{\sigma^2} 
\left[
\left(\overline{v}_{local}-\overline{v} \right)^2 + 
\left(\sigma_{local}-\sigma\right)^2
\right].
\end{equation}
The test statistic, $\Delta$, is the mean value of $\delta_i$
averaged over all galaxies in the sample.

We computed $\Delta$ for our mock clusters using the same galaxies that
were considered when calculating velocity dispersions in Section~5
(i.e. after $3\sigma$-clipping or the application of the den Hartog \&
Katgert algorithm). The results are illustrated in Figure~10 as a
series of plots for the 12 clusters used as examples throughout this
paper. Each galaxy is represented by a circle with radius proportional
to $\exp({\delta_i})$. Individual galaxies with discordant radial
velocities occur in all fields. However, in a number of cases these
discordant galaxies form distinct clumps, indicating the presence of
coherent substructure. We determine the significance of the
subclustering signal using Monte-Carlo simulations. For each cluster we
create 1000 Monte-Carlo realizations by randomly reshuffling radial
velocities, keeping the galaxy positions, and therefore the local
neighbour lists and radial profiles, fixed. The probability that the
observed value of $\Delta$ reflects real substructure is given by one
minus the fraction of these Monte-Carlo simulations that have a value
of $\Delta$ in excess of the observed value. The percentage of clusters
with detected substructure at various confidence levels is given in
Table~1. These percentages are a little lower than the values found by
both Dressler \& Shectman \shortcite{dreshe88} and Bird
\shortcite{bird94}. Since the number of galaxies in our mock clusters
is similar to that in the observational samples, this small discrepancy
could perhaps reflect a bias towards exceptionally rich clusters in the
observational samples, since the degree of substructure is expected to
depend on cluster mass (Lacey \& Cole 1993).

\begin{table}
\caption{Confidence levels of Dressler \& Shectman test}
\begin{tabular}{@{}lcccc}
Confidence level & 100 \% & 99\% & 95\% & 90\% \\ [0.3cm]
Simulated Clusters & 12.7\% & 20.1\% & 31.6\% & 40.1\% \\
Bird (1994) & 24.0\% & 24.0\% & 44.0\% & 52.0\% \\
\end{tabular}
\end{table}

In Figure~11 we plot cluster velocity dispersions, estimated in two
different ways, against the value of the Dressler-Schectman statistic
for each cluster. In the left panel the (2D) velocity dispersions were
obtained using all galaxies within 1.5\hmin\Mpc\ and after applying our
standard $3\sigma$ clipping procedure (n.b. we use the same sample of
galaxies to compute both $\Delta$ and $\sigma_{v,los}$); in the right
hand panel the (2D) dispersions were obtained using den~Hartog \&
Katgert's (1996) interloper removal algorithm. There is a strong
correlation between $\sigma_{v,2D}$ and $\Delta$ in the left hand
panel, confirming our earlier conclusion from Figure~9 that much of the
high $\sigma_v$ tail in the $n(>\sigma_{v,los})$ distribution is due to
artificially large values produced by substructure. The correlation
largely disappears if the den~Hartog-Katgert algorithm is used,
although as we noted above this algorithm tends to slightly
underestimate all velocity dispersions. Our analysis confirms Bird's
(1995) view that cluster masses can be severely overestimated if
substructure is not carefully removed.

\begin{figure}
\centering
\centerline{\epsfysize=9.5truecm
\figinsert{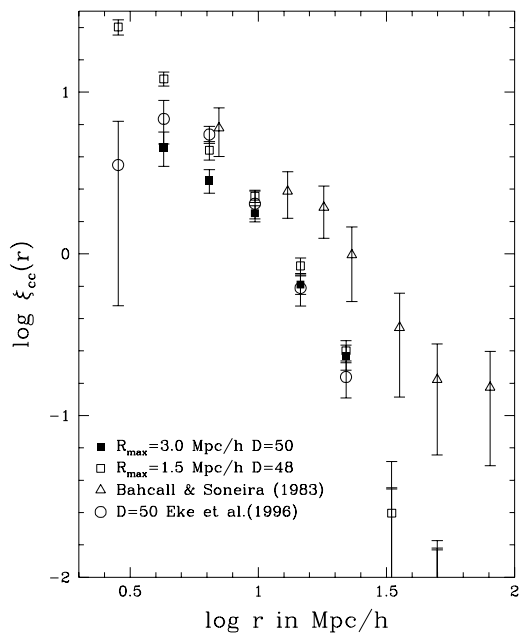}{Figure 12}}
\caption{The two-point correlation function of clusters in redshift
space.  The open triangles show the estimate for $R\ge 1$ Abell
clusters by Bahcall \& Soneira (1983). The open circles show the
correlation function for clusters found in 3D with a mean inter-cluster
separation of $50\hmin\Mpc$. The solid and open squares show results
for clusters identified in 2D, with the poorer of a pair separated by
less than $3\hmin\Mpc$ (solid squares) or $1.5 \hmin\Mpc$ (open
squares) omitted from the sample.}
\end{figure}

\section{The cluster-cluster correlation function} 

The amplitude of the cluster-cluster correlation function, $\xi_{cc}$,
has been a subject of debate for several years, following the original
claim by Bahcall \& Soneira (1983) of a large clustering length,
$r_0\approx 25 \hmin\Mpc$, for R$\geq$1 Abell clusters (where $r_0$ is
defined by $\xi_{cc}(r_0)=1$.) Subsequent redeterminations for Abell
samples have given similar, if somewhat smaller, values \cite{postm92},
but estimates from automated cluster catalogues give much smaller
results, ($r_0\approx 11-14 \hmin\Mpc$) \cite{efst92,nich92,dalton94}.
The latter are closer to (although slightly higher than) the
predictions of the standard CDM model \cite{wfde,croft94,eke96a}.

One of the explanations that has been put forward for the unexpectedly
large value of $r_0$ in the Abell sample is an artificial enhancement
of the correlation function due to projection effects
\cite{suth88,dekel89,efst92}. The apparent richness of sub-threshold
groups projected close to the line-of-sight to a rich cluster may be
overestimated, and the groups promoted to the R$\geq$1 class. This
would introduce spurious close pairs into the sample leading to
artificially high estimates of $\xi_{cc}$ and to distortions in the
correlation function measured as a function of the separation along the
line-of-sight. Sutherland (1988) and Sutherland \& Efstathiou (1992)
argued that these distortions are present in the Abell cluster sample
and, by applying an empirical correction for this effect, derived a low
value of $r_0\approx 13 \hmin\Mpc$ for the Bahcall \& Soneira sample of
Abell clusters.

We can assess the importance of projection effects in the determination
of $\xi_{cc}$ for Abell clusters by comparing estimates for our
simulated samples of clusters identified in 2D and 3D
respectively. Since the cluster correlation function depends on cluster
richness, or equivalently on the mean inter-cluster separation
\cite{bc93,mann93,croft94,eke96a}, we compare mock Abell cluster
samples with a sample of the richest clusters identified in the 3D mass
distribution, with a similar mean inter-cluster separation, $D=50
\hmin\Mpc$. A complication arises from an ambiguity in the way in which
overlapping clusters should be treated. Unfortunately, Abell (1958) did
not specify what action he took when two clusters appeared to overlap
on the sky. We have considered two cases, one in which the poorer of
two clusters that have a separation of less than 3\hmin\Mpc\ is
excluded (i.e. clusters cannot have overlapping Abell radii) and one in
which this restriction is relaxed to 1.5\hmin\Mpc\ (i.e.  a cluster
centre cannot lie within the Abell radius of another cluster.)  Since
this second criterion is less restrictive, 12 per cent more clusters
make it into this sample.

In Figure~12 we compare the redshift-space correlation functions of
clusters identified in 2D (open and solid squares) and 3D (open
circles).  The correlation strength in the 2D sample in which
overlapping Abell radii are allowed exceeds that of the 3D sample on
all scales, but the difference is only large at small pair
separations. As a result, the clustering length of this 2D sample is
only about 10 per cent larger than the value, $r_0=12.2$\hmin\Mpc, for
the sample identified in 3D. This is still significantly smaller than
the clustering length estimated for Abell clusters by Bahcall \&
Soneira (1983) (see Figure~12). Thus, we conclude that projection
effects are not sufficient to enhance the correlation function of
clusters in the standard CDM cosmology to the amplitude estimated by
Bahcall \& Soneira. We note, however, that our mock Abell cluster
sample does not exhibit the strong distortions in the correlation
function along the line-of-sight present in Bahcall \& Soneira's
sample. The enhancement in the correlation function of the 2D sample is
reminiscent of the effect suggested by Sutherland (1988) but the
strength of this effect appears to be weak.

\section{Discussion and conclusions}

We have investigated two sources of uncertainty in the construction and
analysis of cluster catalogues selected from 2D galaxy maps. The first
is the influence of projection effects on the identification and
richness classification of clusters. The second is the influence of
local subclustering on estimates of cluster velocity dispersions. We
have found that both these effects are large and compromise the use of
cluster properties as cosmological diagnostics.

Our conclusions are based on the analysis of mock cluster catalogues.
Although the clusters themselves are identified in 2D galaxy maps using
criteria patterned on those employed in real catalogues, in the
simulations we have access to full 3D spatial and velocity information
for the model galaxies. These were identified using the high peak model
of biased galaxy formation (the peak-background split technique) in an
$\Omega=1$ standard CDM universe with primordial power spectrum
normalized so as to obtain the correct abundance of rich Abell
clusters. The resulting galaxy autocorrelation function is similar to
the observed one in the relevant range of separations
$2<r/\hmin\Mpc<10$.  Clusters were selected from volume-limited galaxy
samples that are complete above the luminosity required to select Abell
clusters over the entire volume. Because the box length in our
simulations is smaller than the effective path length to a moderately
distant Abell cluster, our procedure underestimates the degree of
contamination by distant background galaxies. Our conclusions regarding
the completeness and contamination of Abell's cluster catalogue do not
depend sensitively on the specific cosmological assumptions which we
have chosen for our simulations. On the other hand, our conclusions
regarding the effects of subclustering on estimates of cluster velocity
dispersions, could well be sensitive to our adopted model, particularly
to the assumption that $\Omega=1$, since the degree of substructure in
clusters depends on the value of $\Omega$ (Richstone \etal 1992; Mohr
\etal 1995; Wilson \etal 1996). We intend to investigate this issue in
a subsequent paper.

Most of the clusters identified as rich clusters in our mock catalogues
are associated with at least one large galaxy concentration along the
line-of-sight. However, in almost exactly one third of cases, clusters
classified as having R$\ge 1$ do not correspond to a galaxy
concentration which satisfies the required criterion in
three-dimensions. Instead, the apparent richness of these clusters is
boosted above threshold by the alignment of several smaller clumps
along the line-of-sight. For about half of these spurious R$\ge 1$
clusters, the main concentration along the line-of-sight fails even to
meet the criterion of thirty galaxies required for an R$=0$ cluster. In
a small number of cases (5 per cent) the second largest clump along the
line-of-sight satisfies the R$\ge 1$ richness criterion in its own
right. These clusters should have been included as separate entries in
a complete catalogue. The main source of incompleteness, however, is
clusters that are sufficiently rich, but which are missed because of a
downward fluctuation in the number of background galaxies. Note that
whereas our determination of the background is based on the mean number
of galaxies expected along each line-of-sight, Abell's determination
was based on a local estimate which is not fully explained in his
original paper.  Our results show that Abell's catalogue is neither
homogeneous nor complete above a uniform luminosity limit.

Although our simulations are the first to quantify the importance of
projection effects on Abell's catalogue, the existence of such effects
has been suspected for a long time. For this reason, it has often been
thought that more homogeneous and complete catalogues could be
constructed by selecting clusters according to the X-ray luminosity of
their intra-cluster medium. Because the X-ray emission is centrally
concentrated, projection effects are likely to be much less severe in
this case. Our simulations give partial support to this view. Although
they do not follow the evolution of the X-ray emitting gas, `X-ray
luminosities' can be assigned to clusters in a plausible way that is
consistent with the results of N-body/hydrodynamic simulations. We find
that although many of the spurious optical identifications are avoided
by applying an X-ray luminosity criterion, projection effects are not
completely eliminated. For example, in 16 per cent of cases, the second
most luminous cluster projected along the line-of-sight to another
cluster is bright enough to have been detected in the survey of Briel
\& Henry (1993) even though, in practice, it is unclear whether these
secondary peaks would have been identified as separate clusters.

The abundance of galaxy clusters is a fundamental cosmological
diagnostic which has been used, in various forms, to determine the
amplitude of mass fluctuations (Frenk \etal 1990, Evrard 1989, White
\etal 1993, Viana \& Liddle 1996, Eke \etal 1996b), to rule out
specific cosmological models (Bahcall \& Cen 1993; Lubin \etal 1996),
or to provide an estimator of $\Omega$ (Eke \etal 1996b). Our analysis
of projection effects shows that if the abundance is characterized by
the distribution of velocity dispersion (or by the mass distribution
inferred from it), then results based on Abell's cluster catalogue are
extremely unreliable. Abell's cluster catalogue is so strongly
contaminated that all estimators of velocity dispersion, even elaborate
ones like that of den~Hartog \& Katgert (1996), produce spuriously
large values. For example, a standard $3\sigma$ clipping procedure
applied only to galaxies close to the cluster centre overestimates the
true abundance of clusters with $\sigma_{v,los}$=1500\,\kms\ by more
than an order of magnitude. Tests that rely on the high mass tail of
the distribution derived from optical data are therefore particularly
unreliable.  Since subclustering always tends to inflate the estimated
velocity dispersion, low estimates of $\sigma_{v,los}$ are generally
more reliable than large ones. For example, the estimator mentioned
above leads to a reasonable correlation between estimated and true
values for $\sigma_{v,los}\lsim$ 850 \,\kms, but to no correlation at
all beyond that. As a result, the {\it median} velocity dispersion of
the cluster distribution is reasonably well determined from the data
(White \etal 1993).

Our simulations indicate that increasing the number of galaxies used to
estimate dispersions will not, in general, produce much improvement.
(Dispersions for the most massive of our simulated clusters were
typically measured using over 100 galaxies.) This is because the source
of the problem is the presence of genuine substructure which is not
eliminated by including more galaxies. A better way to improve upon
estimates based on optical data is to use new cluster catalogues such
as the APM catalogue which, by virtue of assuming a smaller search
radius than Abell's, already eliminate much of the contamination.  Even
in this case, however, tests such as those presented here are necessary
to assess the extent of any remaining biases present in the data.

\subsection*{Acknowledgments}
We wish to thank Roland den~Hartog for the use of his interloper
removal routine. MPvH was supported by HCM Grant ERBCHBGCT930454 of the
European Union. This work was also supported by a PPARC rolling grant
for Extragalactic Astronomy and Cosmology at Durham. MPvH acknowledges
the hospitality of the Kapteyn Institute (Groningen). CSF acknowledges
receipt of a PPARC Senior Fellowship.

\end{document}